\documentclass[9pt,twocolumn,twoside]{opticajnl}
\journal{optica} % use for journal or Optica Open submissions
\pdfoutput=1

\setboolean{shortarticle}{true}
\usepackage{verbatim}  % add comment
\usepackage{siunitx}  % use unit
\usepackage{upgreek}
\usepackage{lineno}  % add linenumbers
\newcommand{\suppl}[1]{{\color{blue}{#1}}}

\title{Rigorous and efficient diffraction modeling between arbitrary planes by angular spectrum rearrangement}

\author[1,$\dagger$]{Yiwen Hu}
\author[1,$\dagger$,*]{Xin Liu}
\author[2]{Shi Feng}
\author[1]{Xu Liu}
\author[1,3,4,*]{Xiang Hao}

\affil[1]{College of Optical Science and Engineering, Zhejiang University, Hangzhou 310027, China}
\affil[2]{Department of Pathology, The First Affiliated Hospital, School of Medicine, Zhejiang University, Hangzhou 310003, China}
\affil[3]{Jiaxing Key Laboratory of Photonic Sensing \& Intelligent Imaging, Jiaxing 314000, China}
\affil[4]{Intelligent Optics \& Photonics Research Center, Jiaxing Research Institute Zhejiang University, Jiaxing 314000, China}
\affil[$\dagger$]{Equal contributors}
\affil[*]{Corresponding authors: haox@zju.edu.cn, liuxin2018@zju.edu.cn}

\begin{abstract}
In computational optics, numerical modeling of diffraction between arbitrary planes offers unparalleled flexibility. However, existing methods suffer from the trade-off between computational accuracy and efficiency. To resolve this dilemma, we present a novel approach that rigorously and efficiently models wave propagation between two arbitrary planes. This is achieved by rearranging the angular spectrum of the source field, coupled with linear algebraic computations. Notably, our method achieves comparable computational efficiency to the control method for both scalar and vectorial diffraction modeling, while eliminating nearly all numerical errors. Furthermore, we selectively merge the angular spectrum to further enhance the efficiency at the expense of precision in a controlled manner. Thereafter, the time consumption is reduced to at most 3\% of that before merging.
\end{abstract}

\setboolean{displaycopyright}{false} % Do not include copyright or licensing information in submission.

\begin{document}

\maketitle

% \section{Introduction}
% current challenges (come straight to the point)
The simulation of wave propagation between arbitrary planes has greatly enhanced the versatility of computational optics, yielding important benefits in diverse fields such as microscopy~\cite{Kim:2014:OptExpress:Vectorial}, holography~\cite{Matsushima:2008:Appl.Opt.,Kozacki:2011:Appl.Opt.,Chang:2014:Opt.Express}, and optical tweezers~\cite{Cai:2020:OptExpress:Rapid}. When dealing with diffraction modeling between parallel planes, several algorithms that combine the angular spectrum method and fast Fourier transform (FFT) prove useful in achieving both high computational accuracy and efficiency~\cite{Leutenegger:2006:Opt.Express,Goodman:2017:FourierOptics,Wei2023ModelingOffaxisDiffraction}. However, it becomes more challenging for the non-parallel planes, as the (spatial-) frequency coordinate cannot be directly linked with the desired spatial counterpart using Fourier transform. This limitation imposes constraints on the flexibility of computational optics.

% existing solutions and limitations
Recently, several solutions have been developed, which can be roughly categorized into two groups. The first category combines conventional FFT with angular spectrum interpolation~\cite{Matsushima:2003:JOSAA:Fast,Zhang:2016:ApplOpt:Propagation,Cai:2019:OptCommun:Direct}, allowing for comparable efficiency to that of the parallel case. However, the transform between two uniform frequency domains introduces interpolation errors and fails in large angles due to the additional Jacobian~\cite{Matsushima:2003:JOSAA:Fast}. Alternatively, the second category involves directly calculating the diffractive field using nonuniform FFT algorithms without explicit interpolation~\cite{Chang:2014:Opt.Express,Xiao:2016:JOSAA}, partially warranting computational accuracy. However, it is often tens of times slower compared to conventional FFT and still suffers from numerical errors due to inherent interpolation~\cite{Greengard:2004:SIAMreview}.

% a summary of our solution, methodologies, and results
To resolve this dilemma, in this letter, we present a novel approach for wave propagation modeling between two arbitrary planes. Our method involves a reorganization of the angular spectrum from a uniform frequency domain to a nonuniform one, obviating interpolation. Subsequently, a matrix-based Fourier transform is applied to enable flexible but efficient computations. Additionally, our method also allows a trade-off between precision and efficiency, providing a solution for flexible optimization of performance. We validated our approach both in scalar and in vectorial diffraction modeling scenarios.

Here, we simulated the scalar and the vectorial diffraction with the scalar angular spectrum method (SASM)~\cite{Goodman:2017:FourierOptics} and its vectorial extension (VASM)~\cite{Richards:1959:ProcRSocA:Electromagnetic,Gu:2000:AdvOptImgTheo,Novotny:2012:NanoOptics}, respectively.
While SASM and VASM differ in the treatment of light polarization and are preferred for optical systems with different numerical apertures (NAs), they share the same fundamental interpretation of diffraction---the propagation of the angular spectrum.

\begin{figure}[t]
  \centering
  \includegraphics{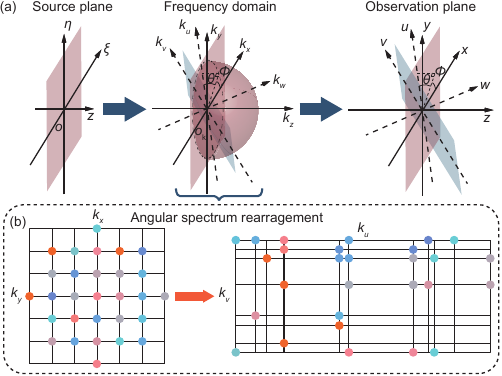}
  \vspace{-6pt}
  \caption{Wave propagation between two arbitrary planes. (a) Coordinate transformation, where the $uv$ plane can be viewed as the $xy$ plane that undergoes two successive rotations. The first rotation involves it rotating around the $z$-axis by an angle of $\phi$, and the second one involves the resulting one rotating around the $v$-axis by an angle of $\theta$. (b) Principles of our method.}
  \label{fig:Fig1_principles}
  \vspace{-6pt}
\end{figure}

The diffraction modeling process, as illustrated in Fig.~\ref{fig:Fig1_principles}(a), comprises three distinct steps. First, the field defined on the source plane $\left(\xi, \eta\right)$ undergoes a transformation into the frequency domain $\mathbf{k}_\mathrm{r} = \left(k_x,k_y,k_z\right)^\mathrm{T}$, yielding the angular spectrum ($\mathrm{T}$ represents transpose). Next, the angular spectrum is projected onto a new coordinate system $\mathbf{k}_\mathrm{t}=\left(k_u,k_v,k_w\right)^\mathrm{T}$, depending on the intersection angle $\left(\theta, \phi\right)$ between the two planes. Herein, $\theta$ and $\phi$ denote the polar and the azimuthal angles, respectively, within the spherical coordinate system. Last, the diffractive field on the observation plane $\left(u, v\right)$ is computed as the Fourier transform of the angular spectrum. Note that the transformation from the source field to the frequency domain is different in SASM and VASM, as detailed in \suppl{Supplement~1}, Sections~ 1--2.

The geometric relation between $\mathbf{k}_\mathrm{r}$ and $\mathbf{k}_\mathrm{t}$ follows $\mathbf{k}_\mathrm{t} = \mathbf{R} \mathbf{k}_\mathrm{r}$, where $\mathbf{R}$ is the rotation matrix [\suppl{Supplement~1}, Eq.~(S4)]. Here we reasonably assume that the source $\xi\eta$, the observation $uv$, and the $k_{x}k_{y}$ planes are all uniformly sampled. As $k_z = \sqrt {k^2 - k_x^2 - k_y^2} $, where $k = 2\uppi n/ \lambda_0$ is the wavenumber in medium with the refractive index $n$ and the vacuum wavelength $\lambda_0$, $k_z$ is nonuniformly sampled, indicating that the angular spectrum in $\mathbf{k}_\mathrm{t}$ space is nonuniformly distributed when the source and the observation planes are not parallel. The major catch is to accurately and efficiently calculate the diffractive field from the nonuniformly distributed angular spectrum. Notably, $\mathbf{k}_\mathrm{t}$ may include many duplicated sampling points along the $k_u$ and $k_v$ axes. By rearranging the duplicated ones to the same coordinate points, a new coordinate system can be established [Fig.~\ref{fig:Fig1_principles}(b)]. This process is termed angular spectrum rearrangement.

To calculate the diffractive field on $\left(u,v\right)$ plane from the angular spectrum in $\mathbf{k}_\mathrm{t}$ space, we employ the matrix triple product (MTP) as a flexible but efficient two-dimensional Fourier transform~\cite{Jurling:2018:JOSAA:Techniques,Liu:2022:OptLasersEng:Fast}, which can handle the case of nonuniform sampling without interpolation and enables arbitrary selection of the sampling intervals and areas without zero-padding. This process can be described by
\begin{equation}
  E \left(u,v\right) = \varOmega\left(v, k_v\right) F\left(k_u, k_v\right) \varOmega\left(u, k_u\right),
\end{equation}
where $E$ and $F$ are the optical fields on the observation plane and the corresponding angular spectrum, respectively. Notably, $F$ is a sparse matrix when the observation plane is tilted. Such a matrix retains the potential to be further compressed. As a result, the memory consumption can be reduced by matrix transformations in subsequent arithmetic processing. $\varOmega\left(u, k_u\right) = \exp \left(iK_u^\mathrm{T} U \right)$ and $\varOmega\left(v, k_v\right) = \exp \left(iV^\mathrm{T} K_v \right)$, where $K_u$, $K_v$, $U$, and $V$ are the coordinates, represented by row vectors, in the frequency and spatial domain, respectively. Assisting with the MTP algorithm, our approach demonstrates the capacity to accurately and efficiently calculate the diffractive field from the nonuniformly distributed angular spectrum.

To establish the efficacy of our method, we conducted a comparative analysis of the numerical results obtained from our approach and a control (Ctrl) method that entails combining the FFT with angular spectrum interpolation~\cite{Zhang:2016:ApplOpt:Propagation,Cai:2019:OptCommun:Direct}. In Ctrl method, we replaced the vanilla FFT with a generalized variant known as the chirp-Z transform to overcome the sampling constraint between the spatial and frequency domains~\cite{Leutenegger:2006:Opt.Express,Jurling:2018:JOSAA:Techniques,Hu:2020:LightSciAppl:Efficient}. The ground truth (GT) is derived from a naive point-by-point integration method~\cite{Jurling:2018:JOSAA:Techniques}. The light wavelength utilized in our simulations is \qty{785}{\nm}. All simulations were executed by MATLAB on a personal computer equipped with an Intel i5 10400F CPU and evaluated with \verb|timeit| function.

% \section{Diffraction modeling}
% \subsection{Scalar diffraction}
\begin{figure}[t]
  \centering
  \includegraphics{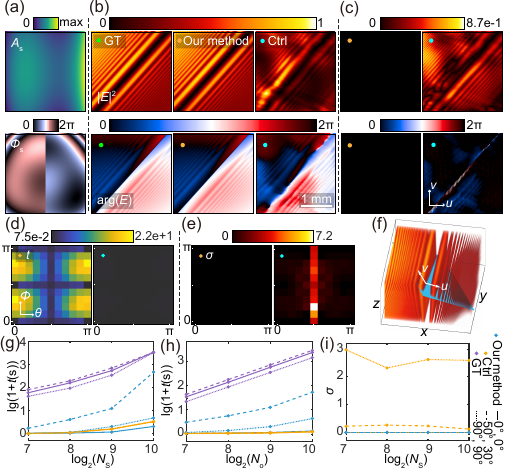}
  \vspace{-6pt}
  \caption{Simulation results of scalar diffraction. (a) The amplitude $A_\mathrm{s}$ and the phase $\varPhi_\mathrm{s}$ distribution on the source plane. (b) The calculated intensity and the phase of the diffractive fields of GT (left column), our method (middle column), and the Ctrl one (right column), respectively. Note that a common linear phase term, induced by the intersection angle, is removed for visualization. (c) The deviations of the results between the two methods and GT, which are computed by direct subtraction. (d) and (e) are the time consumption and the error of the two methods as a function of intersection angles. The error function is defined by \suppl{Supplement~1}, Eq.~(S24). (f) The volumetric intensity distribution of the diffractive field, where the observation plane is colored in blue. (g) and (h) are the time consumptions as a function of the samples on the source and the observation planes, denoted by $N_\mathrm{s}$ and $N_\mathrm{o}$, respectively. (i) The errors of the two methods as a function of $N_\mathrm{s}$.}
  \label{fig:Fig2_scalar_diffraction}
  \vspace{-6pt}
\end{figure}

\begin{figure*}[t]
  \centering
  \includegraphics{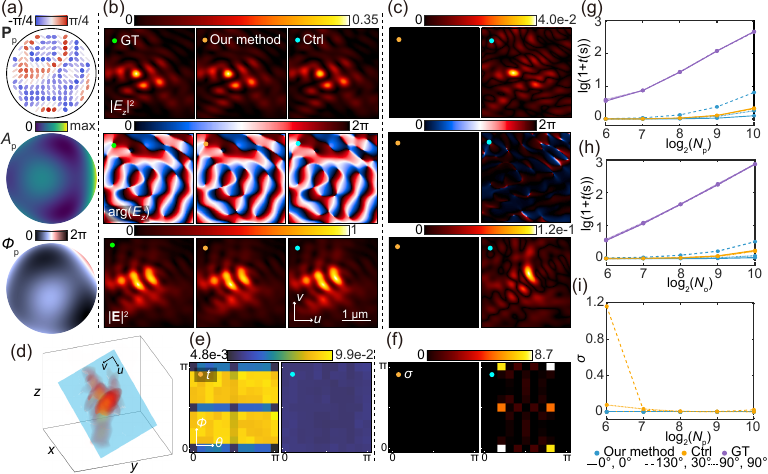}
  \vspace{-6pt}
  \caption{Simulation results of vectorial diffraction. (a) The polarization $\mathbf{P}_\mathrm{p}$, the amplitude $A_\mathrm{p}$, and the phase $\varPhi_\mathrm{p}$ distribution on the pupil plane. The pseudo color for polarization visualization denotes ellipticity. (b) The calculated intensity and the phase of the $z$-polarized component, and the total intensity, of GT (left column), our method (middle column), and the Ctrl one (right column), respectively. The counterparts of $x$- and $y$-polarized components are available in \suppl{Supplement~1}, Section~4A. (c) The deviations of the results between the two methods and GT. (d) The volumetric intensity distribution of the diffractive field, where the observation plane is colored in blue. (e) and (f) are the time consumption and the error of the two methods as a function of angles. (g) and (h) are the time consumptions as a function of the samples on the pupil and the observation planes, denoted by $N_\mathrm{p}$ and $N_\mathrm{o}$, respectively. (i) The errors of the two methods as a function of $N_\mathrm{p}$.}
  \label{fig:Fig3_vectorial_diffraction}
  \vspace{-6pt}
\end{figure*}

We first demonstrated our method on scalar diffraction. The field on the source plane is set to be slightly random to avoid any computational ambiguity and then superposed with a 0-$\uppi$ binary phase to produce some noticeable features in the diffractive field [Fig.~\ref{fig:Fig2_scalar_diffraction}(a)]. The dimensions of the source and the observation planes are $\qty{6.39}{\mm} \times \qty{6.39}{\mm}$ and $\qty{2.56}{\mm} \times \qty{2.56}{\mm}$, respectively. The propagation distance is \qty{50}{\mm}.
% \qty{6387.5}{\um}  \qty{2555}{\um}

We investigated the diffractive field on the observation plane with a randomly selected angle pair $\left(\theta, \phi\right) = \left(\qty{50}{\degree}, \qty{30}{\degree}\right)$ [Fig.~\ref{fig:Fig2_scalar_diffraction}(b)]. The position of the observation plane and the volumetric intensity distribution obtained through the integration method, as a reference, are depicted in Fig.~\ref{fig:Fig2_scalar_diffraction}(f). Both the source and the observation planes are sampled with $512 \times 512$ points. In contrast to the Ctrl one, our method exhibits almost identical results since erroneous points, due to interpolation, are not introduced [Fig.~\ref{fig:Fig2_scalar_diffraction}(c)]. We further quantified the computational accuracy and efficiency at different angles. As illustrated in Fig.~\ref{fig:Fig2_scalar_diffraction}(d), the time consumption of our method is sensitive to angles, reaching the minimum when $\theta$ and $\phi$ are equal to \qty{0}{\degree}, \qty{90}{\degree}, or \qty{180}{\degree}. This is attributed to the tremendous number of duplicated elements of $k_u$ and $k_v$ at these angles, which leaves us the space to maneuver a tighter angular spectrum matrix by rearrangement. Apart from these angles, the duplicated elements are much fewer, resulting in a looser matrix and longer computation time. Nevertheless, our method demonstrates the ability to exactly derive results at arbitrary angles, as depicted in Fig.~\ref{fig:Fig2_scalar_diffraction}(e). In contrast, although the Ctrl one exhibits relatively less and consistent time consumption due to the same samples in $\left(k_u, k_v\right)$ as in $\left(k_x, k_y\right)$, it introduces unacceptable errors with a peak of 7.2 at $\theta = \qty{90}{\degree}$. We are aware that the errors are induced not only by interpolation but also by Jacobian determinant (\suppl{Supplement~1}, Section~1C).

The samples also impact computational efficiency, i.e., more samples slow down the computation. In detail, the integration method is the slowest among the methods [Fig.~\ref{fig:Fig2_scalar_diffraction}(g) and (h)]. Its computational efficiency is slightly affected by the intersection angle, as it determines the sampling requirement in the frequency domain. Meanwhile, the angle also determines if our method can be faster than the Ctrl one, aligning with the findings depicted in Fig.~\ref{fig:Fig2_scalar_diffraction}(d). Fig.~\ref{fig:Fig2_scalar_diffraction}(i) presents the relation between the samples of the source plane and the errors, indicating the ability of our method to consistently yield exact results. However, the Ctrl method suffers from fatal errors when the observation plane is tilted, especially in the orthogonal case due to the Jacobian. To further validate our method, we conducted diffraction modeling of two additional representative source fields: one with a Gaussian amplitude and a 0-$\uppi$ binary phase, and another with a Gaussian amplitude and a thin-lens phase with a circular aperture (\suppl{Supplement~1}, Section 3). Remarkably, all results presented here are reserved.

% \subsection{Vectorial diffraction}

We further demonstrated our method on vectorial diffraction, which is more effective in calculating the tightly focused field formed by a high-NA objective lens (e.g., $\mathrm{NA} = 1.35$ and $n = 1.406$). Similarly, the incident field on the entrance pupil of the objective lens includes slightly random polarization, amplitude, and phase, to represent a general situation [Fig.~\ref{fig:Fig3_vectorial_diffraction}(a)].

The optical field on the observation plane is calculated at $\left(\theta,\phi \right) = \left(\qty{130}{\degree},\qty{30}{\degree} \right)$ [Fig.~\ref{fig:Fig3_vectorial_diffraction}(b)], with the location in the volumetric intensity distribution depicted in Fig.~\ref{fig:Fig3_vectorial_diffraction}(d). Note that here only the $z$-polarized component is visualized since it is a representative feature of a high-NA optical system. The samples on the pupil and the observation planes ($\qty{3.1}{\um} \times \qty{3.1}{\um}$) are $128 \times 128$ and $100 \times 100$, respectively. Similar to the scalar one, our method yields exact results, regardless of the angle [Figs.~\ref{fig:Fig3_vectorial_diffraction}(c) and (f)]. In contrast, the Ctrl one exhibits unacceptable errors of 8.7. Besides, it features a symmetric distribution as shown in Fig.~\ref{fig:Fig3_vectorial_diffraction}(f), which arises from the fact that the two observation planes coincide when their $\theta$ are complementary and $\phi$ are $\qty{0}{\degree}$ and $\qty{180}{\degree}$, respectively. Similar to the scalar case, the time consumption of our method is sensitive to angles, unlike the Ctrl one [Fig.~\ref{fig:Fig3_vectorial_diffraction}(e)]. Yet, even in the worst case, the time consumption of our method is less than 0.1 seconds, which satisfies the requirements of most practical applications.

We also evaluated the performance of three methods as a function of the samples, and the trends generally align with that of the scalar diffraction [Fig.~\ref{fig:Fig3_vectorial_diffraction}(g) and (h)]. Meanwhile, the error of the Ctrl method gradually reduces as the samples increase, indicating that interpolation stands as the primary source of error in the vectorial case [Fig.~\ref{fig:Fig3_vectorial_diffraction}(i)]. To further quantify our method, another representative demonstration, in which the incident field carries a vortex phase, is provided in \suppl{Supplement~1}, Section~4B. In addition, we also examined our method in computer-generated holography between non-parallel planes (\suppl{Supplement~1}, Section~5), which is a representative application of diffraction modeling.

As mentioned above, unlike our method, the Ctrl one suffers from errors induced both by interpolation and by Jacobian. To reveal the interpolation influence, we compared the angular spectrums before and after the coordinate transformation, demonstrating that the interpolation gives rise to considerable errors for the Ctrl method. The detailed analysis is available in \suppl{Supplement~1}, Section 6.

\begin{figure}[t]
  \centering
  \includegraphics{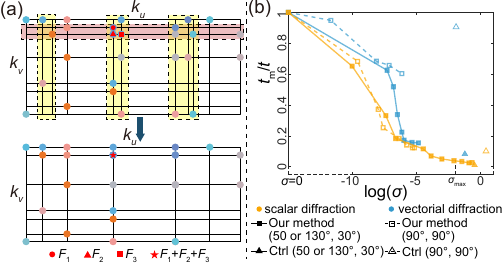}
  \vspace{-6pt}
  \caption{Trade-off between precision and efficiency. (a) Principles of angular spectrum merging. (b) The ratio of the time consumption after merging $t_\mathrm{m}$ to that before merging $t$ as a function of errors. Note that $(\qty{50}{\degree},\qty{30}{\degree})$ and $(\qty{130}{\degree},\qty{30}{\degree})$ are the intersection angles in scalar and vectorial diffraction, respectively.}
  \label{fig:Fig6_tradeoff}
  \vspace{-6pt}
\end{figure}

Although our method guarantees exact results, its computational efficiency is not always promising. To partially solve this problem, we can sacrifice precision in a controlled way for computational efficiency. To accomplish this, we merge the angular spectrums that are located at similar $k_u$ or $k_v$ [Fig.~\ref{fig:Fig6_tradeoff}(a)], which are identified by a threshold. Specifically, these angular spectrums in the red and the yellow boxes are superposed with different weights, respectively, yielding only one element at one $k_u$ or $k_v$ position. This process can be interpreted by approximating angular spectrums, i.e., plane waves, in similar directions to one certain direction. It allows a much tighter angular spectrum matrix, as shown at the bottom of Fig.~\ref{fig:Fig6_tradeoff}(a). By tuning the threshold, we can flexibly control the computational precision and efficiency. The detailed method is referenced in \suppl{Supplement~1}, Section 7.

To verify this strategy, we examined the time consumption of our method as a function of error in both scalar and vectorial diffraction models. Herein, we kept the parameters and the samples consistent with those previously mentioned.  As demonstrated in Fig.~\ref{fig:Fig6_tradeoff}(b), the time consumption of our method diminishes progressively as the error increases, reaching minimum with the minimum samples of angular spectrum. Here we specify the minimum samples of angular spectrum to match those of the source plane, thereby mirroring the Ctrl method. Importantly, the trade-off between precision and efficiency remains consistent at different angles in scalar and vectorial diffraction. For scalar diffraction, as the error increases from 0 to 0.2, the time consumption diminishes to only \qty{3}{\percent} of the original. For vectorial diffraction, when the error increases from 0 to \num{1.4d-5}, the time consumption drops to \qty{15}{\percent} of the original. Assuming the maximum acceptable error $\sigma_\mathrm{max}$ is 0.01, our method can manage to fulfill this condition by weighing the time consumption and error. In contrast, the Ctrl method, whether applied to scalar or vectorial diffraction, consistently fails to bring down the error below the threshold at different angles, even though its time consumption is comparable to our minimum time consumption. Thus, our strategy effectively achieves a notable acceleration within an acceptable error margin. Furthermore, we compared errors and time consumption of the two methods as a function of the samples of the angular spectrum, yielding similar outcomes (\suppl{Supplement~1}, Section 7A and 7B).

% \section{Conclusion}
In conclusion, our study has demonstrated accurate and efficient diffraction modeling between arbitrary planes. It proves applicable for both scalar and vectorial diffraction models. Combined with the MTP algorithm, our approach overcomes the numerical limitations and exhibits high efficiency for observation planes that are parallel or orthogonal to the source plane. For the more general cases, we have also presented a mitigation strategy that allows for a flexible trade-off between precision and efficiency. Furthermore, with the potential for leveraging advanced parallel computing technologies, our method provides access to further acceleration. Our development represents an important step towards flexible, accurate, and efficient computational optics.

\begin{backmatter}
  \bmsection{Funding} Fundamental Research Funds for the Central Universities (2022QZJH29); National Key Research and Development Program of China (2022YFB3206000); "Leading Goose" R\&D Program of Zhejiang (2022C01077); National Natural Science Foundation of China (92050115); Natural Science Foundation of Zhejiang Province (LZ21F050003, LQ20H160048).

  \bmsection{Disclosures} The authors declare no conflicts of interest.

  \bmsection{Data availability} Data may be available from the authors upon reasonable request.

  \bmsection{Supplemental document}
  See \suppl{Supplement~1} for supporting content.

\end{backmatter}

% Bibliography
\bibliography{references}

\begin{thebibliography}{10}
\newcommand{\enquote}[1]{``#1''}

\bibitem{Kim:2014:OptExpress:Vectorial}
J.~Kim, T.~Li, Y.~Wang, and X.~Zhang, \enquote{Vectorial point spread function and optical transfer function in oblique plane imaging,} {\protect\JournalTitle{Optics Express}} \textbf{22}, 11140--11151 (2014).

\bibitem{Matsushima:2008:Appl.Opt.}
K.~Matsushima, \enquote{Formulation of the rotational transformation of wave fields and their application to digital holography,} {\protect\JournalTitle{Applied Optics}} \textbf{47}, D110--D116 (2008).

\bibitem{Kozacki:2011:Appl.Opt.}
T.~Kozacki, \enquote{Holographic display with tilted spatial light modulator,} {\protect\JournalTitle{Applied Optics}} \textbf{50}, 3579--3588 (2011).

\bibitem{Chang:2014:Opt.Express}
C.~Chang, J.~Xia, J.~Wu, W.~Lei, Y.~Xie, M.~Kang, and Q.~Zhang, \enquote{Scaled diffraction calculation between tilted planes using nonuniform fast fourier transform,} {\protect\JournalTitle{Optics Express}} \textbf{22}, 17331--17340 (2014).

\bibitem{Cai:2020:OptExpress:Rapid}
Y.~Cai, S.~Yan, Z.~Wang, R.~Li, Y.~Liang, Y.~Zhou, X.~Li, X.~Yu, M.~Lei, and B.~Yao, \enquote{Rapid tilted-plane gerchberg-saxton algorithm for holographic optical tweezers,} {\protect\JournalTitle{Optics Express}} \textbf{28}, 12729--12739 (2020).

\bibitem{Leutenegger:2006:Opt.Express}
M.~Leutenegger, R.~Rao, R.~Leitgeb, and T.~Lasser, \enquote{Fast focus field calculations,} {\protect\JournalTitle{Optics Express}} \textbf{14}, 11277--11291 (2006).

\bibitem{Goodman:2017:FourierOptics}
J.~W. Goodman, \emph{Introduction to Fourier optics} (W. H. Freeman and Company, 2017), 4th ed.

\bibitem{Wei2023ModelingOffaxisDiffraction}
H.~Wei, X.~Liu, X.~Hao, E.~Y. Lam, and Y.~Peng, \enquote{Modeling off-axis diffraction with the least-sampling angular spectrum method,} {\protect\JournalTitle{Optica}} \textbf{10}, 959--962 (2023).

\bibitem{Matsushima:2003:JOSAA:Fast}
K.~Matsushima, H.~Schimmel, and F.~Wyrowski, \enquote{Fast calculation method for optical diffraction on tilted planes by use of the angular spectrum of plane waves,} {\protect\JournalTitle{Journal of the Optical Society of America A}} \textbf{20}, 1755--1762 (2003).

\bibitem{Zhang:2016:ApplOpt:Propagation}
S.~Zhang, D.~Asoubar, C.~Hellmann, and F.~Wyrowski, \enquote{Propagation of electromagnetic fields between non-parallel planes: a fully vectorial formulation and an efficient implementation,} {\protect\JournalTitle{Applied Optics}} \textbf{55}, 529--38 (2016).

\bibitem{Cai:2019:OptCommun:Direct}
Y.~Cai, Z.~Wang, Y.~Liang, F.~Ren, B.~Yao, M.~Lei, and S.~Yan, \enquote{Direct calculation of tightly focused field in an arbitrary plane,} {\protect\JournalTitle{Optics Communications}} \textbf{450}, 329--334 (2019).

\bibitem{Xiao:2016:JOSAA}
Y.~Xiao, X.~Tang, Y.~Qin, H.~Peng, W.~Wang, and L.~Zhong, \enquote{Nonuniform fast fourier transform method for numerical diffraction simulation on tilted planes,} {\protect\JournalTitle{Journal of the Optical Society of America A}} \textbf{33}, 2027--2033 (2016).

\bibitem{Greengard:2004:SIAMreview}
L.~Greengard and J.-Y. Lee, \enquote{Accelerating the nonuniform fast fourier transform,} {\protect\JournalTitle{SIAM review}} \textbf{46}, 443--454 (2004).

\bibitem{Richards:1959:ProcRSocA:Electromagnetic}
B.~Richards and E.~Wolf, \enquote{Electromagnetic diffraction in optical systems, ii. structure of the image field in an aplanatic system,} {\protect\JournalTitle{Proceedings of the Royal Society of London. Series A}} \textbf{253}, 358--379 (1959).

\bibitem{Gu:2000:AdvOptImgTheo}
M.~Gu, \emph{Advanced Optical Imaging Theory}, Springer Series in Optical Sciences (2000).

\bibitem{Novotny:2012:NanoOptics}
L.~Novotny and B.~Hecht, \emph{Principles of nano-optics} (Cambridge university press, 2012).

\bibitem{Jurling:2018:JOSAA:Techniques}
A.~S. Jurling, M.~D. Bergkoetter, and J.~R. Fienup, \enquote{Techniques for arbitrary sampling in two-dimensional fourier transforms,} {\protect\JournalTitle{Journal of the Optical Society of America A}} \textbf{35}, 1784--1796 (2018).

\bibitem{Liu:2022:OptLasersEng:Fast}
X.~Liu, Y.~Hu, S.~Tu, C.~Kuang, X.~Liu, and X.~Hao, \enquote{Fast generation of arbitrary optical focus array,} {\protect\JournalTitle{Optics and Lasers in Engineering}} \textbf{162}, 107405 (2023).

\bibitem{Hu:2020:LightSciAppl:Efficient}
Y.~Hu, Z.~Wang, X.~Wang, S.~Ji, C.~Zhang, J.~Li, W.~Zhu, D.~Wu, and J.~Chu, \enquote{Efficient full-path optical calculation of scalar and vector diffraction using the bluestein method,} {\protect\JournalTitle{Light: Science and Applications}} \textbf{9}, 119 (2020).

\end{thebibliography}

% Full bibliography added automatically for Optics Letters submissions; the following line will simply be ignored if submitting to other journals.
% Note that this extra page will not count against page length
\bibliographyfullrefs{references}

\end{document}